\pdfoutput=1
\documentclass[12pt]{article}
\newcommand{\comm}[1]{}

\newcommand{\newc}{\newcommand}\newc{\beq}{\begin{equation}}\newc{\eeq}{\end{equation}}
\newc{\bal}{\begin{align}}\newc{\eal}{\end{align}}
\newc{\ba}{\begin{eqnarray}}\newc{\ea}{\end{eqnarray}}\newc{\bea}{\begin{eqnarray*}}
\newc{\eea}{\end{eqnarray*}}\newc{\alp}{\alpha}
\newc{\eps}{\epsilon}\newc{\vph}{\varphi}
\newc{\vhp}{\varphi}\newc{\tx}{\tilde {x}}

\usepackage{graphicx,rotating,colortbl}
\usepackage{jcappub}\usepackage{amsmath}\usepackage{amssymb}\usepackage{braket}
\usepackage{graphicx}\usepackage{hyperref}
\usepackage{mathrsfs}\usepackage{subcaption}
\usepackage{empheq}\usepackage{appendix}\usepackage{natbib}\usepackage[dvipsnames]{xcolor}
\usepackage{float}\usepackage[top=1in, bottom=0.4in, left=0.8in, right=0.8in, includefoot]{geometry} 
\usepackage[autostyle]{csquotes}
\usepackage{color,soul}
\usepackage{mathtools}%\coloneqq

\definecolor{DarkViolet}{RGB}{148,0,211}
\hypersetup{colorlinks,bookmarksopen,
bookmarksnumbered,
citecolor={DarkViolet},
linkcolor={DarkViolet},
urlcolor={DarkViolet},
pdfstartview=FitH}

\usepackage{eulervm}
\definecolor{DarkBlue}{RGB}{0,0,153}
\newcommand{\Subsection}[1]{\textcolor{black}{\subsection{#1}}}
\newcommand{\Section}[1]{\textcolor{black}{\section{#1}}}

\usepackage{scalerel}
\usepackage{tikz}
\usetikzlibrary{svg.path}
\definecolor{orcidlogocol}{HTML}{A6CE39}
\tikzset{
  orcidlogo/.pic={
    \fill[orcidlogocol] svg{M256,128c0,70.7-57.3,128-128,128C57.3,256,0,198.7,0,128C0,57.3,57.3,0,128,0C198.7,0,256,57.3,256,128z};
    \fill[white] svg{M86.3,186.2H70.9V79.1h15.4v48.4V186.2z}
                 svg{M108.9,79.1h41.6c39.6,0,57,28.3,57,53.6c0,27.5-21.5,53.6-56.8,53.6h-41.8V79.1z M124.3,172.4h24.5c34.9,0,42.9-26.5,42.9-39.7c0-21.5-13.7-39.7-43.7-39.7h-23.7V172.4z}
                 svg{M88.7,56.8c0,5.5-4.5,10.1-10.1,10.1c-5.6,0-10.1-4.6-10.1-10.1c0-5.6,4.5-10.1,10.1-10.1C84.2,46.7,88.7,51.3,88.7,56.8z};}}
\newcommand\orcid[1]{\href{https://orcid.org/#1}{\mbox{\scalerel*{
\begin{tikzpicture}[yscale=-1,transform shape]
\pic{orcidlogo};
\end{tikzpicture}
}{|}}}}
\graphicspath{{./Figs/}}

\title{General parametrization of higher-dimensional black holes and its application to Einstein-Lovelock theory}

\author[a,b]{Roman A.~Konoplya \orcid{0000-0003-1343-9584},}
\author[a]{Thomas D.~Pappas \orcid{0000-0003-2186-357X},}
\author{and}
\author[a]{Zden\v{e}k Stuchl\'ik}

\emailAdd{roman.konoplya@gmail.com}
\emailAdd{thomas.pappas@physics.slu.cz}
\emailAdd{zdenek.stuchlik@physics.slu.cz}

\affiliation[a]{Research Centre for Theoretical Physics and Astrophysics, Institute of Physics, Silesian University in Opava, Bezručovo nám. 13, CZ-746 01 Opava, Czech Republic}

\affiliation[b]{Peoples Friendship University of Russia (RUDN University),
6 Miklukho-Maklaya Street, Moscow 117198, Russian Federation}

\abstract{Here we have developed the general parametrization for spherically symmetric and asymptotically flat black-hole spacetimes in an arbitrary metric theory of gravity.  The parametrization is similar in spirit to the parametrized post-Newtonian (PPN) approximation, but valid in the whole space outside the event horizon, including the near horizon region. This generalizes the continued-fraction expansion method in terms of a compact radial coordinate suggested by Rezzolla and Zhidenko [Phys.Rev.D 90 8, 084009 (2014)] for the four-dimensional case. As the first application of our higher-dimensional parametrization we have approximated black-hole solutions of the Einstein-Lovelock theory in various dimensions. This allows one to write down the black-hole solution which depends on many parameters (coupling constants in front of higher curvature terms) in a  very compact analytic form, which depends only upon a few parameters of the parametrization. The approximate metric deviates from the exact (but extremely cumbersome) expressions by fractions of one percent even at the first order of the continued-fraction expansion, which is confirmed here by computation of observable quantities, such as quasinormal modes of the black hole.}

\begin{document}
\maketitle
\vspace{0.8cm}

\Section{Introduction\label{sec:intro}}

\noindent Black holes in theories with higher-curvature corrections play an important role in high-energy physics, from the tentative form of quantum corrections to gravity in the low-energy limit of string theory \cite{Boulware:1985wk,Wiltshire:1985us,Kanti:1995vq} to the description of strongly coupled quantum systems within the AdS/CFT correspondence \cite{Grozdanov:2016fkt,Brigante:2007nu}. One of the most promising approaches is given by the Einstein-Lovelock theory, a generalization of the Einsteinian theory, which is the most general metric theory of gravity yielding conserved second-order equations of motion in an arbitrary number of spacetime dimensions $D$ \cite{Lovelock:1971yv,Lovelock:1972vz}. When one is limited by the quadratic correction in curvature, the corresponding limit of the Lovelock theory reproduces the Gauss-Bonnet combination, the first black-hole solution for which was obtained by Boulware and Deser \cite{Boulware:1985wk}. In four dimensional spacetimes, the Gauss-Bonnet term is a pure divergence, and thus the corresponding field equations remain unaltered. In five- and six- dimensional spacetimes, the Einstein-Gauss-Bonnet action is the most general, while for higher dimensions, higher-order corrections in curvature must be used for consistency and generality. Each correction term of the $m$th power in curvature in the infinite Lovelock series contains a dimensional coupling constant $\tilde {\alpha}_{m}$ which is divided by some power of the radius $r_0$ of the event horizon,
\begin{equation}
mth~Lovelock~term \sim \tilde {\alpha}_{m} r_{0}^{-2 m +2}
\end{equation}
so that the smaller black hole is, the more terms of the Lovelock theory are important. While for sufficiently large black holes the first, quadratic, Gauss-Bonnet correction is sufficient, for smaller black holes one need to take more and more Lovelock terms into consideration. At the same time, even the cubic correction makes the black-hole metric function very cumbersome, because it cannot be expressed in a general closed form for the whole set of parameters, but includes finding of roots of some algebraic equations for determination of the metric at a given set of parameters. Then, the description of the properties and observable quantities for such small black holes in the Einstein-Lovelock theory of high order would be an almost never-ending task: analysis of black-hole behavior depending on even several parameters requires very large resources and provides big room for interpretations, while adding more parameters simply makes the problem unfeasible and the role of each Lovelock term uncertain.

Here we suggest the way to solve the above problem, that is, we develop a formalism which, being much wider and not linked to any particular form of the metric theory, allows one to describe a cumbersome analytical or numerical black-hole solution, depending on a large number of parameters of a theory, in a compact analytical form which depends only upon a few new parameters of the parametrization. This is done via the introduction of the general parametrization of any spherically symmetric and asymptotically flat black-hole spacetime by using the continued-fraction expansion in terms of a compact radial coordinate. The general ansatz for the parametrization is designed in such a way that the prefactors determine the asymptotic behavior, while the terms in the continued fraction series are fixed by the behavior near the event horizon. This way, the parametrization is valid not only near the black hole or only in the far region, but in the whole space outside the black hole. This idea was first applied to the four-dimensional case by Rezolla and Zhidenko \cite{Rezzolla2014}, and extended to axially symmetric 4D black holes by Konoplya, Rezzolla and Zhidenko \cite{Konoplya:2016jvv} and was effectively applied in a number of recent works to analysis of four-dimensional black holes in various theories of gravity \cite{Younsi:2016azx,Kokkotas:2017zwt,Kokkotas:2017ymc,Konoplya:2018arm,Nampalliwar:2018iru,Konoplya:2019,Konoplya2019}.

In this work we generalize the aforementioned approach to an arbitrary number of dimensions. Therefore, the general parametrization developed here can be applied to any spherically symmetric and asymptotically flat black-hole metric independently on the character of a theory of gravity under consideration. After developing of the general parametrization we apply it to the Einstein-Lovelock black holes and show that, for example, the fifth-order Einstein-Lovelock black hole which depends upon five coupling constants can be represented in a very compact form which depends only on two parameters. Further we illustrate the effectiveness of this approach by calculating the quasinormal modes for this system.

The further extension of the method to incorporate the parametrization of rotating higher-dimensional (HD) black holes in full generality is a highly nontrivial task that will be the subject of a future work. Nevertheless, as a first step toward this direction, we have investigated here the case of slowly rotating HD solutions with a single rotation parameter. Under these assumptions the deviation of the metric from spherical symmetry is encoded in a single off-diagonal metric component. We have developed a parametrization for the off-diagonal metric function and applied it to the case of slowly rotating Lovelock black holes where we find excellent agreement already in the first order in the approximation.

The paper is organized as follows. In Sec.~\ref{sec:4dCFA}  we briefly review the Rezzolla-Zhidenko parametrization for  four-dimensional static black holes, while in Sec.~\ref{sec:HDCFA} the HD generalization of the parametrization has been developed. In Sec.~\ref{sec:sl_rot} we introduce the HD parametrization in the case of slow rotation around a single axis.  The remainder of the article is dedicated to the application of the generalized parametrization. We start in Sec.~\ref{sec:Lovebh} by giving basic information about higher-dimensional black holes in the Einstein-Lovelock theory. Section~\ref{sec:LoveHDCFA} is devoted to the application of the parametrization for the approximation of the Einstein-Lovelock black-hole metric for various dimensions and Lovelock curvature orders and Sec.~\ref{sec:HDCFASR} deals with the approximation of slowly rotating Lovelock black holes. Section~\ref{sec:QNMs} is dedicated to the study of the quasinormal modes of Einstein-Lovelock black holes in the presence of a large number of higher-curvature terms and finally, in Sec.~\ref{sec:Conclusions} we summarize the obtained results and mention some open problems.

\Section{The continued-fraction parametrization}

\Subsection{Parametrization of four dimensional black-hole metrics\label{sec:4dCFA}}

\noindent In this section we shall briefly review the method introduced in \cite{Rezzolla2014} for the parametrization of black-hole metrics in terms of a continued-fraction expansion. This will help to set the stage for its generalization in the following section. Consider a spherically symmetric and asymptotically flat black-hole solution of the Einstein equations described by the line-element ansatz
\beq
ds^2=-f(r)dt^2+\frac{dr^2}{h(r)}+r^2 \left(d\theta^2+ \sin^2\theta\, d \phi^2 \right)\,.
\eeq
The real positive roots of the equation $f(r)=0$ correspond to the radii of the horizons present in the spacetime. We symbolize the radius of the outer event horizon of the black hole by $r_0$ and restrict our analysis for the remainder of this article on events satisfying the condition $r \geqslant r_0$. We may perform a coordinate transformation and introduce a dimensionless compact coordinate via
\beq
x(r)\equiv 1-\frac{r_0}{r}\,,
\label{4dxdef}
\eeq
that ranges from $x=0$ for $r=r_0$ up to $x=1$ in the limit of $r \rightarrow \infty$. Then, in terms of the compact coordinate and a set of constant parameters, we may reexpress the metric functions through a set of parametrization equations as
\begin{eqnarray}
f(r)&=&x A(x)\,,\nonumber\\
\frac{f(r)}{h(r)}&=& B(x)^2\,,
\label{metAB}
\end{eqnarray}
where the two new functions on the right-hand side (r.h.s.) of the above equations are defined as
\begin{eqnarray}
A(x)&\equiv &1-\eps\, (1-x)+(a_0-\eps)(1-x)^2+{\tilde  A}(x)(1-x)^3\,,\nonumber\\
B(x)&\equiv &1+b_0\,(1-x)+{\tilde  B}(x)(1-x)^2\,,
\label{defAxBx}
\end{eqnarray}
and the functions ${\tilde  A}(x)$ and ${\tilde  B}(x)$ are constructed by means of continued-fraction expansions as follows:
\beq
{\tilde  A}(x)=\frac{a_1}{\displaystyle 1+\frac{\displaystyle
    a_2\,x}{\displaystyle 1+\frac{\displaystyle a_3\,x}{\displaystyle
      1+\ldots}}}\,,\,\,\,
      {\tilde  B}(x)=\frac{b_1}{\displaystyle 1+\frac{\displaystyle
    b_2\,x}{\displaystyle 1+\frac{\displaystyle b_3\,x}{\displaystyle
      1+\ldots}}}\,.
      \label{TildeAB}
\eeq
Note that the parameters of Eqs.~\eqref{defAxBx} can be divided into two sets depending on the way they are determined. The first set consists of the triad of the \enquote{asymptotic} parameters $(\eps,a_0,b_0)$ that are specified upon comparing terms of the same order in the expansions of Eqs.~\eqref{metAB} at spatial infinity. In the second set we have the continued-fraction parameters $(a_1,a_2,\cdots,b_1,b_2,\cdots)$ that are determined by the corresponding expansions of Eqs.~\eqref{metAB} in the vicinity of the event horizon.

In the limit of an infinite number of expansion terms, the parametrization~\eqref{metAB} reproduces a given metric function exactly for all $r \geqslant  r_0 $. At the same time, the functional dependence of the parametrization on $x$ via continued fractions provides impressive convergence properties and thus, in practice, only a small number of terms is required in order to yield a highly accurate continued-fraction approximation (CFA) of a given metric. The truncation of the series at a finite expansion order $m$ is easily achieved by fixing $a_m=b_m=0$ and thus higher-order parameters play no role in the analysis. It is usually the case that the accuracy of the approximation is increased by one order of magnitude with each order in the continued-fraction expansion and typically the first few terms suffice for the description of observable quantities with an absolute relative error in the range of fractions of one percent. We would also like to emphasize that one of the most attractive features of the CFA scheme is that it provides a very accurate description for the metric function not only close to the black hole horizon or in the far field but for all values of the radial coordinate $r \in [r_0,\infty)$.

In principle, when working in four dimensions, there are observational constraints associated with the post-Newtonian expansion in the far region that force the values of the asymptotic parameters $a_0$ and $b_0$ to be $\mathcal{O}(10^{-4})$ \cite{Rezzolla2014,Will2014a}. It is then common practice in 4D analyses to fix \emph{$a$ priori} $a_0=b_0=0$ in Eqs.~\eqref{defAxBx} but for now we shall retain them as free parameters since we follow a more general approach of the CFA method for illustrative reasons.

In the far-field region, the metric functions for an arbitrary asymptotically flat black hole may be expanded as a power series of falloff terms ($\sim 1/r^{n},\,\,n \geqslant1$) in the following way:
\beq
f(r)=1+\sum_{n=1}^{\infty} \frac{f_n}{r^n}\,,\,\,\,h(r)=1+\sum_{n=1}^{\infty} \frac{h_n}{r^n}\,,
\label{FFexp}
\eeq
where the set of the expansion coefficients ($f_n,h_n$) can in principle be determined as functions of the free parameters of the system upon direct substitution of the expansion~\eqref{FFexp} into the field equations of the theory at hand. The coefficient of the $r^{-1}$ term of the expansion of $f(r)$ is then associated with the asymptotic black-hole mass $h_1=-2M$ and the coefficient of $r^{-2}$ with the charge $Q^2$ of the solution.

Upon expanding both sides of Eqs.~\eqref{metAB} in the far region ($r \rightarrow \infty$\,,\,\,$x \rightarrow 1$) and comparing terms of the same order, one finds that in full generality, the asymptotic parameters are given in terms of the coefficients of the asymptotic expansions as
\beq
\eps=-\left(1+\frac{f_1}{r_0} \right)\,,\,\,\,a_0=\frac{f_2}{r_0^2}\,,\,\,\,b_0=\frac{f_1-h_1}{2\, r_0} \,,
\eeq
or equivalently in terms of $M$ and $Q^2$
\beq
\eps=\frac{2M}{r_0}-1\,,\,\,\,a_0=\frac{Q^2}{r_0^2}\,,\,\,\,b_0=-\frac{2M+h_1}{2\, r_0} \,.
\label{4dasympt}
\eeq
By the form of these equations it is clear that the parameter $\eps$ measures the deviation of the black hole event-horizon radius $r_0$ from the Schwarzschild radius $r_{Sch}=2M$. Also, the fact that the parameter $a_0$ is proportional to the charge of the black hole complies with the observational constraints that limit its values to $\mathcal{O}(10^{-4})$ \cite{Rezzolla2014} for astrophysically relevant configurations as we have already mentioned. We point out that $a_0$ has also been found to be proportional to the charge in the case of scalarized Einstein-Maxwell black holes \cite{Konoplya2019}.

For the pure Schwarzschild black hole one has the identifications $r_0 = 2M\,,Q^2=0\,,f_1=h_1 = -2M$ and $f_i = 0\,, h_i = 0\,,\forall i \geqslant 2$ and so, we see that all of the asymptotic parameters vanish in this limit. As for the continued-fraction parameters, in the Schwarzschild limit we have $a_1 \rightarrow 0\,,\,\, b_1 \rightarrow 0\,,$ which essentially truncates the CFA at the zeroth order. It is in this sense then that the Schwarzschild metric plays the role of the \enquote{reference} metric around which this approximation method has been built and the CFA parameters encapsulate the deviations of a given metric from it.

When applying the CFA method, either in order to obtain an analytic representation for a numerical solution or in the case of a cumbersome analytic solution in order to have a more compact expression to perform calculations with, we need to determine the parameters ($M\,,Q^2\,,h_1$) in order to have the CFA parameters. The former parameters of the metric functions can be easily computed in both cases by isolating the coefficients of $r^{-1}$ (for $f(r)$ and $h(r)$) and $r^{-2}$ (for $f(r)$) in their far-field expansions.

\Subsection{The extension of the method for higher-dimensional metrics\label{sec:HDCFA}}

\noindent Gravitational theories with more than three spatial dimensions in the framework of general relativity\footnote{Historically, the idea to consider extra spatial dimesnions for the first time is credited to G. Nordstr\"om in 1912 \cite{Ravndal2004}.} (GR) can be traced back to the first attempts toward a unified theory of gravity and electromagnetism by Kaluza and Klein \cite{Kaluza:1921tu,Klein:1926tv}. The first black-hole solution of the Einstein equations in $D$ dimensions has been derived by F.R.Tangherlini in the early $1960$s and constitutes the natural HD generalization of the Schwarzschild metric along with the possible inclusion of charge and cosmological constant terms \cite{Tangherlini1963}. A couple of decades later, a significant resurgence of interest in HD gravity and its black-hole solutions emerged as a byproduct of the advent of string theory and since then the subject has been exhaustively investigated in a plethora of contexts (see for example \cite{Kanti:2004nr,Emparan2008,Maartens2010} and references therein).

Here, we are interested in black-hole solutions to the $D-$dimensional Einstein equations that are spherically symmetric, asymptotically flat and stem from an arbitrary metric theory of gravity. In general, we may write the metric ansatz for such a black hole as
\beq
ds^2= -f(r)\,dt^2+\frac{1}{h(r)}dr^2+r^2 d\Omega_{D-2}^2\,,
\label{HDmetricAnsatz}
\eeq
where $d\Omega_{D-2}^2$ is the line element on the unit $(D-2)-$sphere. The extra dimensions enter the line element in the form of $n=D-4$ extra angular coordinates labeled as $\theta_n$ with $n \geqslant 1$ in terms of which we have
\beq
d\Omega_{D-2}^2= d \theta_n^2+\sin^2\theta_n \left[d\theta_{n-1}^2+\sin^2\theta_{n-1}\left[\cdots+\sin^2 \theta_{1} \left[ d \theta^2 +\sin^2\theta\, d \phi^2  \right] \cdots\right] \right] \,.
\eeq
The metric functions $f(r)$ and $h(r)$ depend on the radial coordinate as well as on the dimensionality of spacetime. The invariance of the metric under time translations implies that the radii of the horizons correspond to the real positive roots of the equation $f(r)=0$ and the outer black-hole event horizon will be once again denoted by $r_0$.

In order to extend the formalism of the previous section to the case of HD metrics, we introduce a new generalized radial compact coordinate as follows:
\beq
\tx(r) \equiv 1- \left(\frac{r_0}{r} \right)^q= 1- \left(\frac{r_0}{r} \right)^{D-3}\,.
\label{HDxdef}
\eeq
For reasons of text compactness, we have already assigned the appropriate value to the parameter $q$ in the above equation but in order to justify this choice, let us for the moment consider that it is a yet undetermined parameter.

By construction, $\tx(r)$ shares the same asymptotic values with $x(r)$~\eqref{4dxdef} for any $q \in \mathcal{N}$, both close to the horizon ($\lim_{r \rightarrow r_0} \tx(r) =x(0)=0$) and at spatial infinity ($\lim_{r \rightarrow \infty} \tx(r) =\lim_{r \rightarrow \infty}x(r)=1$). To specify the value of $q$ we will turn to the asymptotic expansion of the parametrization equations which have once again the following form:
\begin{eqnarray}
f(r)&=&\tx \left[ 1-\eps\, (1-\tx)+(a_0-\eps)(1-\tx)^2+\frac{a_1}{\displaystyle 1+\frac{\displaystyle
    a_2\,\tx}{\displaystyle 1+\ldots}}(1-\tx)^3 \right]\,,\label{HD_f_param}\\
\frac{f(r)}{h(r)}&=& \left[ 1+b_0\,(1-\tx)+\frac{b_1}{\displaystyle 1+\frac{\displaystyle
    b_2\,\tx}{\displaystyle 1+\ldots}}(1-\tx)^2 \right]^2\,.
\label{HD_foh_param}
\end{eqnarray}
Evidently, when $D=4\, \rightarrow \,q=1$ we have $\tx(r)=x(r)$ and thus the four-dimensional method \cite{Rezzolla2014} is included as a special case in  our more general framework. In the far region, the expansions for the metric functions of an arbitrary asymptotically flat metric in $D$ dimensions have the following form:
\beq
f(r)=1+\sum_{n=1}^{\infty} \frac{f_n}{r^{n(D-3)}}\,,\,\,\,h(r)=1+\sum_{n=1}^{\infty} \frac{h_n}{r^{n(D-3)}}\,.
\label{asympt_gtt_HD}
\eeq
The effective mass of the solution is associated with the coefficient of the lowest-order term of the expansion of $f(r)$ namely $r^{-(D-3)}$ while the effective charge is related to the second term of the series $r^{-2(D-3)}$. Upon substituting Eq.~\eqref{asympt_gtt_HD} in the left-hand side (l.h.s.) of Eq.~\eqref{HD_f_param} and re-expressing the rhs in terms of the  original radial coordinate $r$ via Eq.~\eqref{HDxdef}, we find that at spatial infinity the corresponding lowest-order term on the r.h.s. is proportional to $r^{-q}$ and so we are led to identify $q=D-3$.

Alternatively, based on the following observation, one could have intuitively postulated that $q=D-3$ from the beginning. As we have seen, the four-dimensional parametrization is build around the Schwarzschild metric and the compact coordinate $x(r)$ corresponds to the $g_{tt}(r)$ metric function of the Schwarzschild solution when $r_0=r_{Sch}$. Then, the analogue \enquote{reference metric} upon which one can construct the HD parametrization is the asymptotically flat and uncharged Tangherlini solution \cite{Tangherlini1963} whose $g_{tt}(r)$ metric component is given exactly by~\eqref{HDxdef} for $q=D-3$.

Via the expansions of Eqs.~\eqref{HD_f_param} and~\eqref{HD_foh_param} in the asymptotic region we determine once again the form of the asymptotic parameters (with $f_1 \equiv -\mu$ and $f_2 \equiv \mathcal{Q}^2$) as
\beq
\eps=\frac{\mu}{r_0^{(D-3)}}-1\,,\,\,\,a_0=\frac{\mathcal{Q}^2}{r_0^{2(D-3)}}\,,\,\,\,b_0=-\frac{\mu+h_1}{2\, r_0^{(D-3)}} \,,
\label{HDasympt}
\eeq
that reduce to Eqs.~\eqref{4dasympt} for $D=4$. In Eq.~\eqref{HDasympt}, the mass parameter $\mu$ is related to the black-hole mass $M$ \cite{Myers:1988ze}
\beq
\mu =\frac{16 \pi G_D\, M}{(D-2)\, \Omega_{D-2}}\,,\,\,\,\Omega_{D-2}=\frac{2\,\pi^{\frac{D-1}{2}}}{\Gamma\left(\frac{D-1}{2}\right)}\,,
\label{mu_def}
\eeq
where $G_D$ is the gravitational constant in $D$ dimensions, $\Omega_{D-2}$ is the area of the unit $(D-2)-$sphere, and the asymptotic charge $\mathcal{Q}^2$ is related to the charge of the black hole $Q^2$ via \cite{Frassino2014}
\beq
\mathcal{Q}^2=\frac{8 \pi G_D\, Q^2}{(D-2)(D-3)}\,.
\label{Q2_def}
\eeq
Note that in HD theories there are no observational constraints similarly to the ones imposed on the asymptotic parameters of the 4D-CFA via the PPN formalism. This means that there is no reason to assume \emph{$a$ priori} that $a_0$ and $b_0$ will be negligibly small parameters. Instead, one has to obtain their values by isolating the coefficients of the appropriate asymptotic terms of the metric function $f(r)$ at spatial infinity as we have already discussed in the previous section.

\Subsection{Parametrization of slowly rotating higher-dimensional black holes\label{sec:sl_rot}}

\noindent In four dimensions, the extention of the parametrization to incorporate rotating black hole metrics turned out to be a far-from-trivial task \cite{Rezzolla2014} but nonetheless it has been achieved in \cite{Konoplya:2016jvv}. In the presence of extra dimensions, the problem becomes even more complicated since there are in principle\footnote{The \emph{floor} function of a number $\lfloor a \rfloor$ gives the largest integer $b$ that satisfies $b \leqslant a$.} $\lfloor (D-1)/2 \rfloor$ independent angular-momentum parameters associated with all the possible directions for rotation in the bulk \cite{Myers1986}. To this end, we postpone the comprehensive study and analysis of this extension for a future work and we focus here in the special case of metrics with a single rotation parameter $a$ associated with the angular momentum of a black hole that rotates on a single two-plane that lies on the brane.

The \enquote{reference metric} for our parametrization here will be the Myers-Perry black hole \cite{Myers1986} with a spherical event-horizon topology and a single axis of rotation which is described by the line element
\begin{eqnarray}
ds^{2} & = &
-\left( 1- \frac {\mu }{\Sigma\, r^{D-5}} \right) dt^{2}
-\frac {2a\mu \sin ^{2} \theta }{\Sigma\, r^{D-5}} \, dt \, d\phi
+\frac {\Sigma }{\Delta } \, dr^{2}
+ \Sigma \, d\theta ^{2}
\nonumber \\ & &
+ \left( r^{2} + a^{2} + \frac {a^{2}\mu \sin ^{2} \theta}{\Sigma\, r^{D-5}}
\right) \sin ^{2} \theta \, d\phi ^{2}
+r^{2} \cos ^{2}\theta \, d\Omega _{D-4}^{2}\,,
\label{MyersPerry}
\end{eqnarray}
where
\begin{equation}
\Delta = r^{2} +a^{2} - \frac {\mu }{r^{D-5}}, \qquad
\Sigma = r^{2}+a^{2} \cos ^{2} \theta\,.
\label{MPDelta}
\end{equation}
The mass parameter $\mu$ is related to the mass $M$ of the black hole via Eq.~\eqref{mu_def} while the rotation parameter $a$ is associated to both the angular momentum $J$ and mass $M$ of the black hole via
\begin{equation}
a=\frac{D-2}{2}\,\frac{J}{M}\,,
\end{equation}
and thus it can be interpreted as the angular momentum per unit mass. Once we impose the slow-rotation condition ($J \ll M \rightarrow  a \ll 1 $) on Eq.~\eqref{MyersPerry} by retaining only terms of $\mathcal{O}(a)$ and generalizing the metric functions to arbitrary functions of the radial coordinate we are led to consider the following ansatz for a general slowly rotating HD black hole (see also \cite{Aliev2006,Camanho2015,Adair2020}):
\beq
ds^2= -f(r)\,dt^2+\frac{1}{h(r)}dr^2-2\, \omega(r)r^2 \sin^2\theta dt\,d\phi+r^2\left(d\Omega_{2}^2+\cos^2\theta\, d \Omega^2_{D-4}\right)+\mathcal{O}(\omega^2)\,.
\label{SRHDansatz}
\eeq
The metric function $\omega(r)$ is in principle arbitrary, but it should vanish in the static-limit of the solution ($a=0$) and should also exhibit the appropriate asymptotic profile at spatial infinity. More precisely,
in $D$ dimensions, from the expansion of the off-diagonal metric function for the Meyers-Perry solution one has for $\omega(r)$ \cite{Myers1986}:
\beq
\omega(r) \approx \frac{8 \pi G_D\,J}{\Omega_{(D-2)}\,r^{D-1}}\,,
\label{omegaDasympt}
\eeq
and so for a general solution $\omega(r)$ the lowest-order falloff term should be $\sim r^{-(D-1)}$.

Regarding the radius of the event horizon of the general slowly rotating black hole~\eqref{SRHDansatz}, we point out that the metric~\eqref{MyersPerry}, and consequently~\eqref{SRHDansatz}, does not depend on the coordinates $t$ and $\phi$, and so it is endowed with the Killing vectors $\left( \partial / \partial t \right)^\mu$ and $\left( \partial / \partial \phi \right)^\mu$ respectively. A linear combination of the two defines the following Killing vector field \cite{Carroll:2004st,Visser2007}:
\beq
K^{\mu} \equiv  \left(\frac{\partial}{\partial t} \right)^{\mu}+\Omega_H \left(\frac{\partial}{\partial \phi} \right)^{\mu}\,,
\label{KillingK}
\eeq
where $\Omega_H$ is the \enquote{angular velocity} of the event horizon and is of order $\mathcal{O}(a)$. In stationary and asymptotically flat spacetimes any event horizon is a Killing horizon i.e. a hypersurface where $K^{\mu}$ becomes null \cite{Faraoni:2015ula}. The vanishing of $K^{\mu}K_{\mu}$ on the event horizon, yields the following equation for the determination of the radii of the horizons:
\beq
g_{tt}+2\, \Omega_H\, g_{t \phi}+\Omega_H^2\, g_{\phi \phi}=0 \Rightarrow g_{tt}=0 +\mathcal{O}(a^2)\,.
\eeq
In the last step we used the fact that $\omega(r)$ should be of $\mathcal{O}(a)$ to lowest order in the slow-rotation approximation. Thus we conclude that the radius of the outer event horizon will be the same as in the non-rotating case and equal to $r_0$ as specified by the equation $g_{tt}(r_0)=f(r_0)=0$. In the same spirit with the previous sections, we may rewrite the function $\omega(r)$ in terms of a continued-fraction expansion and the generalized compact coordinate of Eq.~\eqref{HDxdef} as
\beq
\omega(r)\,r^{2} =\omega_0 \left(1-\tilde {x} \right)+\frac{\omega_1}{\displaystyle 1+\frac{\displaystyle
    \omega_2\, \tilde {x}}{\displaystyle 1+\ldots}}(1-\tilde {x})^2 \,.
    \label{OmegaCFA}
\eeq
Note that the presence of $r^2$ on the l.h.s. of the above parametrization equation is of pivotal importance in order for Eq.~\eqref{OmegaCFA} to yield the appropriate asymptotic terms for $\omega(r)$. More precisely, upon rearranging the last equation at spatial infinity ($\tx \approx 1$) we have
\beq
\omega(r)=\frac{\omega_0}{r^{2}}\, \left(\frac{r_0}{r} \right)^{D-3}+\mathcal{O}\left[ \left( 1-\tilde {x} \right)^{2} \right]\,,
\eeq
where in the first term we have used the definition of $\tx$~\eqref{HDxdef}. Then, comparison with Eq.~\eqref{omegaDasympt} reveals that the asymptotic parameter in Eq.~\eqref{OmegaCFA} in the case of the Meyers-Perry \enquote{reference metric} will be given by
\beq
\omega_0=\frac{8 \pi G_D J}{\Omega_{(D-2)}r_0^{D-3}}\,,
\label{w0_ref}
\eeq
which is exactly the value of the rotation parameter $a$ and so $\omega_0$ is expected to be of $\mathcal{O}(a)$ for metrics that do not deviate drastically from the Meyers-Perry solution. The remaining continued-fraction parameters $\omega_m$ with $m \geqslant 1$, will be determined by comparison of the series expansion of Eq.~\eqref{OmegaCFA} in the vicinity of the event horizon $r_0$.

\Section{Applying the general parametrization for Lovelock black holes}

\Subsection{Black-hole solutions in Lovelock gravity\label{sec:Lovebh}}

\noindent In the context of the classical theory of GR the gravity sector of the action consists solely of the Einstein-Hilbert term and the cosmological constant. With such a minimal setup, the corresponding field equations yield solutions that are sufficient to comply with nearly all observations to date including the recent detection of gravitational waves \cite{Collaboration2016,Laurentis2016} and the shadow of the supermassive black hole M87${}^{*}$ \cite{Collaboration2019}.

Deviations from the predictions of GR do emerge in observations at galactic and cosmological scales but this alone is not a sufficient argument in favor of the necessity for the modification of the gravity sector. The reason being that the aforementioned issues might be remedied or at least ameliorated by modifying the energy-momentum tensor in the Einstein equations for example by considering new fundamental fields beyond the Standard Model. On this basis, some might argue that GR is still not facing any serious conflict with current observations.

The by now undeniable predictive power of GR has been repeatedly put to the test for more than a century and passed with flying colors, albeit in energy scales much lower than the Planck scale where quantum gravity effects are expected to become important. It is generally believed that GR is the low-energy limit of a more fundamental theory and modifications of the gravity sector in the action should be taken into account when analyzing strong-gravity phenomena.

The most general pure-gravity extension of GR in any number of dimensions by means of the inclusion of higher-curvature terms in the action that become important at high energy yielding second-order field equations and thus avoiding the emergence of ghosts \cite{Zwiebach:1985uq} has been derived by D. Lovelock in 1971 \cite{Lovelock:1971yv}. Explicitly, in $D=4+n$ dimensions the action for Lovelock gravity is \cite{Cai2003,Takahashi2010}
\beq
S=\int{d^D\,x\sqrt{-g} \sum_{m=0}^{k} c_m \,\mathcal{L}_m}\,,
\label{Action:Lovelock}
\eeq
where $k=\lfloor \frac{D-1}{2} \rfloor$ is the maximum Lovelock order and $c_m$ are arbitrary coupling constants of the theory with dimensions [length]${}^{2m-D}$. The $m$th order term is constructed out of contractions of $m$ powers of the Riemann tensor and is written explicitly as
\beq
\mathcal{L}_m=\frac{1}{2^{m}} \delta^{\mu_1 \nu_1\cdots\mu_m\nu_m}_{\rho_1 \lambda_1\cdots\rho_m\lambda_m}\, R^{\rho_1 \lambda_1}_{\quad\;\;\mu_1 \nu_1} \cdots R^{\rho_m \lambda_m}_{\quad\;\;\mu_m \nu_m}\,,
\eeq
where the generalized totally antisymmetric Kronecker delta is defined via
\beq
\delta^{\mu_1 \mu_2\cdots\mu_m}_{\nu_1 \nu2\cdots\nu_m} \equiv \begin{vmatrix}
\delta^{\mu_1}_{\nu_1} & \delta^{\mu_1}_{\nu_2} & \cdots & \delta^{\mu_1}_{\nu_m}\\
\delta^{\mu_2}_{\nu_1} & \delta^{\mu_2}_{\nu_2} & \cdots & \delta^{\mu_2}_{\nu_m}\\
\vdots & \vdots & \ddots & \vdots\\
\delta^{\mu_m}_{\nu_1} & \delta^{\mu_m}_{\nu_2} & \cdots & \delta^{\mu_m}_{\nu_m}\\
\end{vmatrix}\,.
\eeq
The zeroth- and first-order terms of the Lovelock series~\eqref{Action:Lovelock} correspond to the cosmological constant and Einstein-Hilbert term respectively i.e.
\beq
c_0 \,\mathcal{L}_0 = -2 \Lambda \,,\,\,\,c_1\, \mathcal{L}_1 = \frac{R}{16 \pi G_D} \,.
\label{Lovelock1and2}
\eeq
Note that since for $D=4$ the maximum Lovelock order is $k=1$ we are left with GR as the only theory of the Lovelock family in four dimensions. In this work we are interested in asymptotically flat solutions and so we shall set $c_0=0$ and also fix $c_1=(16 \pi G_D)^{-1}=1$. The second-order Lovelock term contributes nontrivially in HD theories $D>4$ and corresponds to the Gauss-Bonnet invariant
\beq
c_2\, \mathcal{L}_2 = c_2 \left( R^2 -4R_{\mu\nu}R^{\mu\nu}+R_{\mu\nu\rho\sigma}R^{\mu\nu\rho\sigma} \right)\,,
\eeq
while the next higher-order correction to the action, $c_3\, \mathcal{L}_3$ comes into play for $D>6$. In practice, in order to work in the framework of a HD theory we will usually need to make a postulation about the dimensionality $D$ of spacetime and this in turn will naturally truncate the Lovelock series at a finite order. In any case, the resulting metric function will depend in principle on a large number of Lovelock coupling parameters making the analytic description of the solutions quite cumbersome and impractical for analytic computational purposes.

The metric function of spherically symmetric black-hole solutions in Lovelock gravity \cite{Boulware:1985wk,Wheeler:1985nh,Myers:1988ze,Cai2003,Garraffo2008,Frassino2014} for an arbitrary number of dimensions can be obtained as 
\beq
f(r)=1-r^2\,\psi(r)\,.
\label{metric function}
\eeq
The function $\psi(r)$ is a solution of the following algebraic equation that emerges upon substituting Eq.~\eqref{metric function} into the field equations and involves a polynomial of degree $k \equiv \lfloor \frac{D-1}{2} \rfloor$
\beq
W[\psi(r)]=\psi(r)+\sum_{m=2}^{k}\tilde {\alp}_m \psi(r)^m= \frac{\mu}{r^{D-1}}-\frac{\mathcal{Q}^2}{r^{2(D-2)}} \,,
\label{Lovelock_eq}
\eeq
where, the constant of integration $\mu$ is the mass parameter~\eqref{mu_def}, $\mathcal{Q}^2$ is the asymptotic effective charge~\eqref{Q2_def} that appears in the Lovelock equation~\eqref{Lovelock_eq} when the action~\eqref{Action:Lovelock} is supplemented with the inclusion of the Maxwell term, and for brevity we have defined
\beq
\tilde {\alp}_m \equiv c_m \displaystyle\prod_{p=1}^{2m-2} (D-p-2)=c_m\, \frac{(D-3)\,!}{(D-2m-1)\,!} \,.
\eeq

In the simple case of the first nontrivial Lovelock correction to GR, the algebraic equation~\eqref{Lovelock_eq} is quadratic and of the two solutions that emerge for $\psi (r)$ only one  yields a metric function that has a smooth Einstein gravity limit, i.e., the Tangherlini solution can be recovered as $\tilde {\alpha}_2 \rightarrow 0$. We denote this solution by $\psi_{GB}(r)$ and the corresponding metric function describes the Gauss-Bonnet black hole in $D$ dimensions:
\beq
f_{GB}(r)=1-r^2\psi_{GB}(r)=1+\frac{1}{2\, \tilde {\alpha}_2}\, r^2 \left[1- \sqrt{1+4\, \tilde {\alpha}_2 \left( \frac{\mu}{r^{D-1}}- \frac{\mathcal{Q}^2}{r^{2(D-2)}}\right)} \right]\,.
\label{GB_metric}
\eeq

From Eq.~\eqref{metric function} we can readily find that the value for the function $\psi (r)$ at the location of the event horizon is  $\psi (r_0)=r_0^{-2}$. Also, we can solve the Lovelock equation \eqref{Lovelock_eq} at $r=r_0$ in order to express the mass parameter $\mu$ in terms of the radius of the event horizon and the rest of the free parameters of the system as
\beq
\mu= r_0^{D-3}+\frac{\mathcal{Q}^2}{r_0^{D-3}}+\sum_{m=2}^{k}\tilde {\alp}_m \, r_0^{D-2 m-1}\,.
\label{mu}
\eeq
In the above expression, we see that the corrections induced on $\mu$ by the presence of the higher-curvature terms of the Lovelock series involve the Lovelock coupling parameters normalized by some power of the event-horizon radius. Consequently, they can be safely ignored for sufficiently large black holes but on the other hand they become important for small black holes. If we denote the characteristic length-scale of extra dimensions by $L$ then any black hole with $r_0 \gg L$ is effectively four-dimensional and can be described as a HD object only if $r_0 \ll L$. Thus, when considering HD black holes, the Lovelock corrections are indeed important and should be taken into account.

As stability analyses of the Lovelock black holes have revealed \cite{Konoplya2017} the values of $\tilde {\alp}_m$ cannot be arbitrary if we are interested in stable solutions. In the next section, restricting our analysis to values of $\tilde {\alp}_m$ that yield stable black holes, we will employ the HDCFA in order to test the accuracy of the method and obtain compact approximations for the cumbersome expressions of the Lovelock black-hole metric functions.

\Subsection{Compact expressions for the metric of Lovelock black holes\label{sec:LoveHDCFA}}

\noindent In \cite{Konoplya2017}, the stability of the black-hole solutions emerging in Lovelock theory in arbitrary curvature order and dimensionality have been investigated. Furthermore, a numerical code has been made publicly available therein that tests whether a given set of Lovelock parameters corresponds to a stable black hole and if not, which types of instabilities plague the solution.

In this section, for a wide range of number of dimensions i.e. $D \in [5,11]$ spanning maximal Lovelock orders up to $k=5$ we employed the HDCFA to obtain approximations at various orders in the continued-fraction expansion. The sets of the Lovelock parameters that we used have been tested with the aforementioned code and yield stable and asymptotically flat black-hole configurations.

\begin{figure*}[h!]
\includegraphics[width=0.5\linewidth]{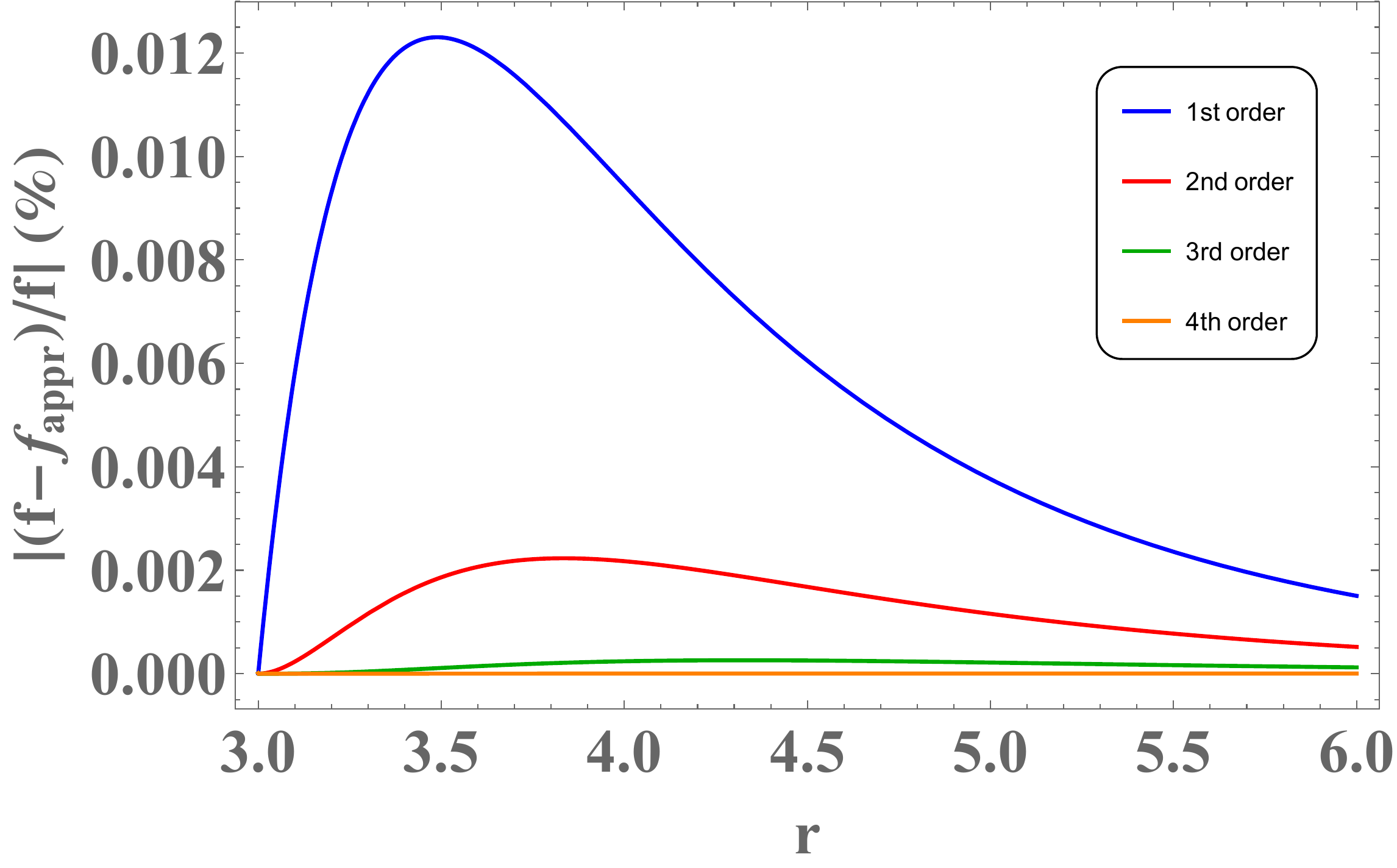}\includegraphics[width=0.5\linewidth]{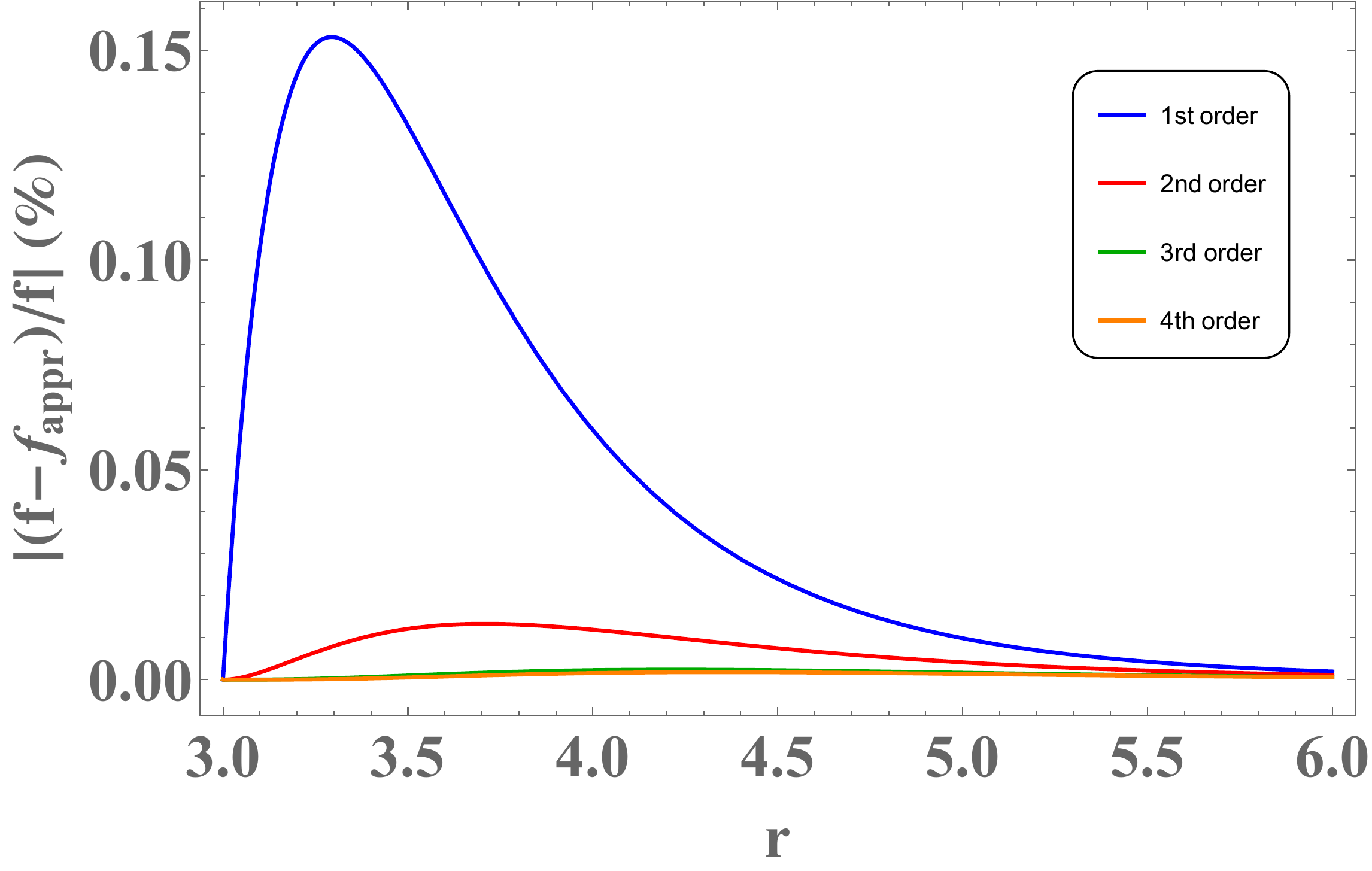}
\includegraphics[width=0.5\linewidth]{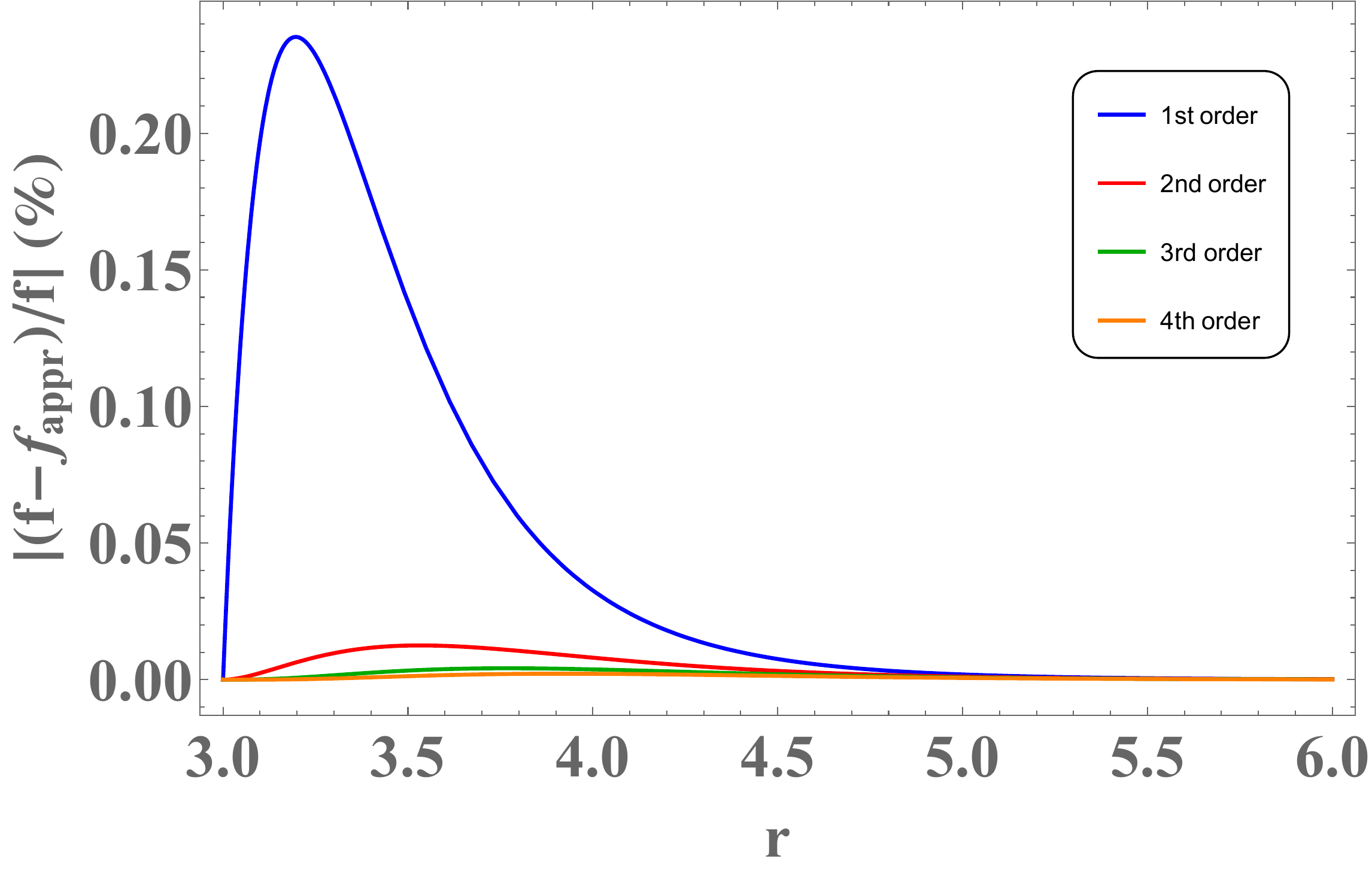}\includegraphics[width=0.5\linewidth]{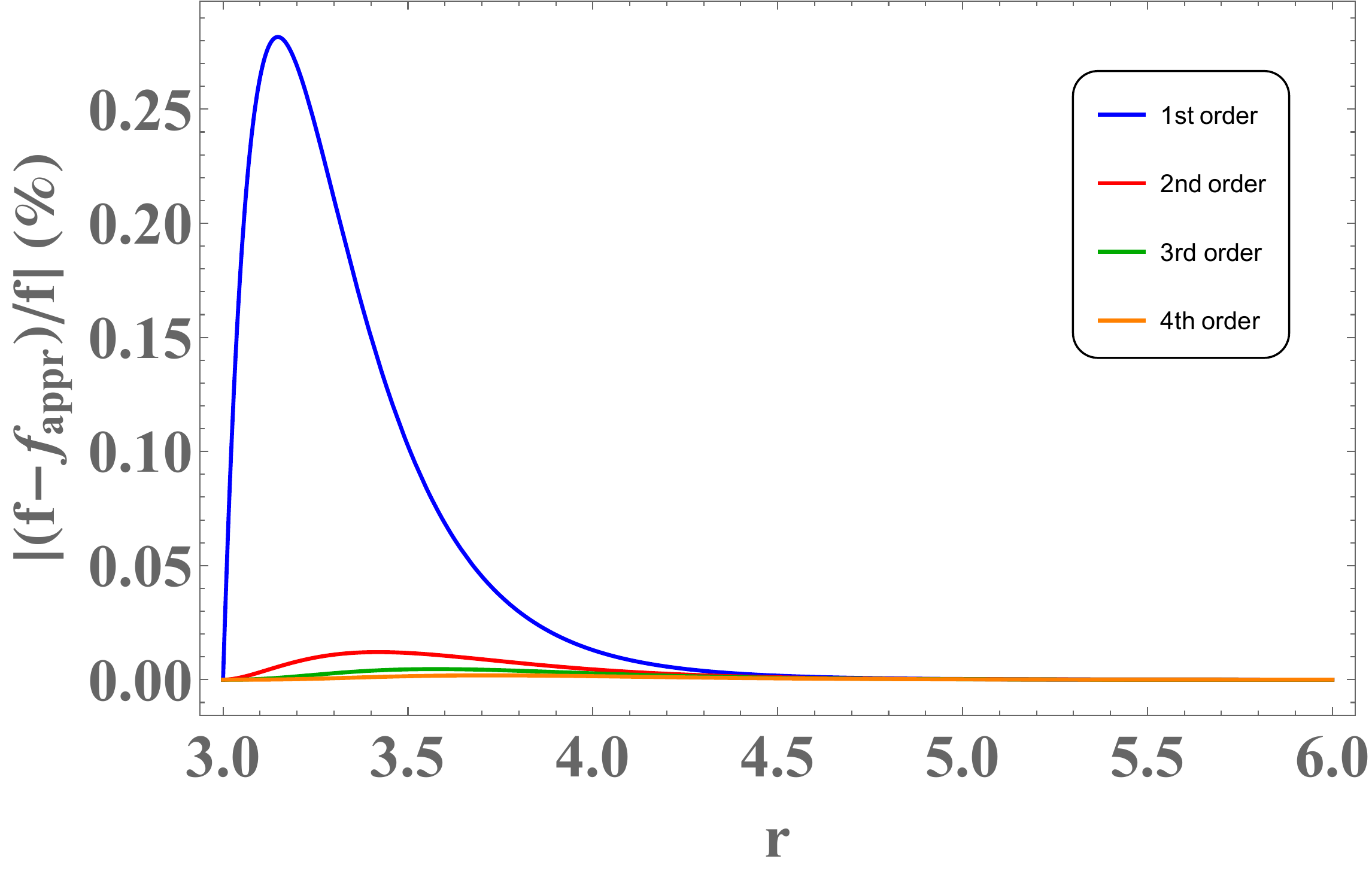}
\caption{The absolute relative error between the exact solutions $f(r)$ and the HDCFA $f_{appr}(r)$ at the first four orders of the approximation expressed in percentages ($\%$), for $D=5$ (top left), $D=7$ (top right), $D=9$ (bottom left) and $D=11$ (bottom right). The values of the parameters are $(r_0=3\,,\mathcal{Q}=\frac{2}{3},\,\tilde {\alp}_0=0\,,\tilde {\alp}_1=1\,,\tilde {\alp}_2=\frac{1}{4}\,,\tilde {\alp}_3=\frac{1}{5}\,,\tilde {\alp}_4=\frac{1}{7}\,,\tilde {\alp}_5=\frac{1}{10})$\,.
} \label{fig:AREs}
\end{figure*}

In Fig.~\ref{fig:AREs} we plot the absolute relative error (ARE) for the metric function at different orders in the HDCFA with respect to the exact expression. The radial profile of the AREs is typical of the CFA scheme where the maximum value (MARE) is located close to the event horizon ($\sim 1.1\, r_0 - 1.3\, r_0$) and asymptotes monotonically to negligible values both as $r \rightarrow r_0$ and $r \rightarrow \infty$.

\begin{table}[h!]
\center{
\begin{tabular}{|c|c|c|c|c|c|c|c|}
 \hline
 HDCFA order
  &  $D=5$ & $D=6$ & $D=7$ & $D=8$ & $D=9$ & $D=10$ & $D=11$ \\
 \hline\hline
  1 & 0.0123 & 0.0894 & 0.1532 & 0.1998 & 0.2353 & 0.2611 & 0.2816 \\
 \hline
  2 &  0.0022 & 0.0122 & 0.0133 & 0.0149 & 0.0125 & 0.0128 & 0.0121 \\
 \hline
  3 &  0.0003 & 0.0029 & 0.0024 & 0.0035 & 0.0042 & 0.0044 & 0.0046 \\
 \hline
   4 &  $\mathcal{O}(10^{-6})$ & 0.0006 & 0.0018 & 0.0020 & 0.0022 & 0.0022 & 0.0019 \\
 \hline
\end{tabular}
\caption{The maximal absolute relative error in percentages ($\%$) between the exact metric function and its approximation for various dimensions ($D$) and orders in the HDCFA. The values of the parameters here are fixed to $(r_0=3\,,\mathcal{Q}=\frac{2}{3},\,\tilde {\alp}_0=0\,,\tilde {\alp}_1=1\,,\tilde {\alp}_2=\frac{1}{4}\,,\tilde {\alp}_3=\frac{1}{5}\,,\tilde {\alp}_4=\frac{1}{7}\,,\tilde {\alp}_5=\frac{1}{10})$\,.}\label{table:MARE}
}
\end{table}

As the elements of Table~\ref{table:MARE} verify, the HDCFA converges since the MAREs become smaller by approximatelly one order of magnitude (at least for the first few orders) as we increase the order of the approximation by one. It is also clear that HDCFA provides an excellent approximation of the exact solutions for Lovelock black holes at different dimensions only with a few parameters. In fact, even at the first order of the approximation we have a MARE for the metric function that is only a small fraction of $1\%$. This means, that the highly complicated exact expressions involving a plethora of Lovelock parameters belong in the family of the so-called \enquote{moderate metrics} \cite{Konoplya2020a} that can be accurately approximated at the first order in HDCFA and thus be described in terms of a simple analytic expression. Recently it has been shown that it may be possible to formulate this concept of a moderate metric using a mathematically strict invariant measure \cite{Suvorov:2020klq}. Possible constraints on the black-hole parametrization coming from experiments related to observations of gravitational waves have been discussed in \cite{Volkel:2020daa}.

By truncating the continued-fraction expansion in Eq.~\eqref{HD_f_param} to the first order we end up with the following analytic form that can describe moderate metrics for any charged and asymptotically flat HD black hole in an arbitrary theory of gravity:
\beq
f_{mod}(r)=1-\left(\eps+1 \right) \left( \frac{r_0}{r}\right)^{D-3}+a_0\left( \frac{r_0}{r}\right)^{2\,(D-3)}+\left( a_1-a_0+\eps \right)\left( \frac{r_0}{r}\right)^{3\,(D-3)}-a_1 \left( \frac{r_0}{r}\right)^{4\,(D-3)}\,,
\label{fHDmod}
\eeq
where $\eps$ and $a_0$ are defined in Eq.~\eqref{HDasympt}.

In the case of Lovelock black holes, $\mu$ is given by Eq.~\eqref{mu} and $a_1$ turns out to have the simple and very compact form
%\beq
%a_1= \frac{2\, \mu}{r_0^{D-3}}-\frac{\mathcal{Q}^2}{r_0^{2(D-3)}}-3+\frac{2+r_0^3\, \psi'(r_0)}{3-D}=2\,\eps -1 -a_0 +\frac{2+r_0^3\, \psi'(r_0)}{3-D}\,,
%\label{anala1}
%\eeq
\begin{eqnarray}
a_1 & = & \frac{2\, \mu}{r_0^{D-3}}-\frac{\mathcal{Q}^2}{r_0^{2(D-3)}}-3+\frac{2+r_0^3\, \psi'(r_0)}{3-D}
\nonumber \\ & = &
2\,\eps -1 -a_0 +\frac{2+r_0^3\, \psi'(r_0)}{3-D}\,,
\label{anala1}
\end{eqnarray}
where we have used $\psi(r_0)=r_0^{-2}$ and $\psi'(r_0)$ is given by
\beq
\psi'(r_0)=\frac{(D-3)\,\mathcal{Q}^2+(1-D)\,\sum_{m=1}^{k} \tilde {\alp}_m\, r_0^{2(k+l-m)} }{\sum_{m=1}^{k} m\, \tilde {\alp}_m\, r_0^{2(k+l-m)+3} }\,,
\label{psiprime}
\eeq
with $k \equiv \lfloor \frac{D-1}{2} \rfloor$ and $l \equiv \lfloor \frac{D-2}{2} \rfloor$ for $D\geqslant 5$.

Thus, for a given fixed set of values for the free parameters $(r_0,\mathcal{Q},\tilde {\alp}_m)$ one obtains a compact and very accurate (MARE at fractions of one percent) analytic expression for the metric function of any asymptotically flat Lovelock black hole. In case an even more accurate approximation is required, we provide a Mathematica\textregistered{} notebook as an ancillary file where one can obtain compact analytic expressions for the metric function for any $D\geqslant 5$ and at any desired order in the HDCFA.

\Subsection{Slowly rotating Lovelock black holes\label{sec:HDCFASR}}

\noindent In order to test the accuracy of the proposed parametrization of Sec.~\ref{sec:sl_rot}, let us consider the metric for a slowly rotating, charged and asymptotically flat Lovelock black hole in D dimensions described by the line element \cite{Camanho2015,Adair2020}
\beq
ds^2= -\left[ 1-r^2\,\psi(r) \right]\,dt^2+\frac{dr^2}{\left[ 1-r^2\,\psi(r) \right]}-2\, a\,\psi(r)\,r^2  \sin^2\theta dt\,d\phi+r^2\left(d\Omega_{2}^2+\cos^2\theta\, d \Omega^2_{D-4}\right) \,,
\label{SRLove}
\eeq
where $a$ is the rotation parameter and $\psi(r)$ is a solution to the Lovelock equation~\eqref{Lovelock_eq}. An inspection of the off-diagonal metric component reveals that the function we need to approximate by means of the parametrization equation~\eqref{OmegaCFA} is
\beq
\omega(r)\,r^{2} =a\, \psi(r)\,r^{2}\,.
\eeq

A comparison of the expansions of the metric functions at spatial infinity and close to the event horizon determines the values of the parameters $\omega_m\,,m \geqslant 0$. Focusing for illustration in the five-dimensional case for which only the Gauss-Bonnet correction term is relevant and thus only the $\tilde {\alp}_2$ Lovelock coupling comes into play \eqref{GB_metric}, the first two parameters of the CFA have the following expressions:
\beq
\omega_0=a\left( 1+\frac{ \mathcal{Q}^2+\tilde {\alpha}_2 r_0^2}{r_0^4}\right)\,,\qquad \omega_1= -a \left(\frac{\mathcal{Q}^2}{r_0^4}+\frac{\tilde {\alp}_2}{r_0^2}+\frac{r_0^2}{2 \tilde {\alp}_2}-\frac{\sqrt{\left(2 \tilde {\alp}_2+r_0^2\right)^2}}{2 \tilde {\alp}_2} \right) \,,
\eeq
and are indeed proportional to the rotation parameter $a$ as expected. This ensures that in the static limit we recover the non-rotating Lovelock black hole. Once we have a parametrization for $a\, \psi(r)\,r^{2}$ to the desired order in the continued-fraction expansion, it is then straightforward to obtain the approximation for the full off-diagonal component of Eq.~\eqref{SRLove} by multiplying the result with $-2\,\sin^2\theta$.

\begin{figure*}[h!]
\includegraphics[width=0.5\linewidth]{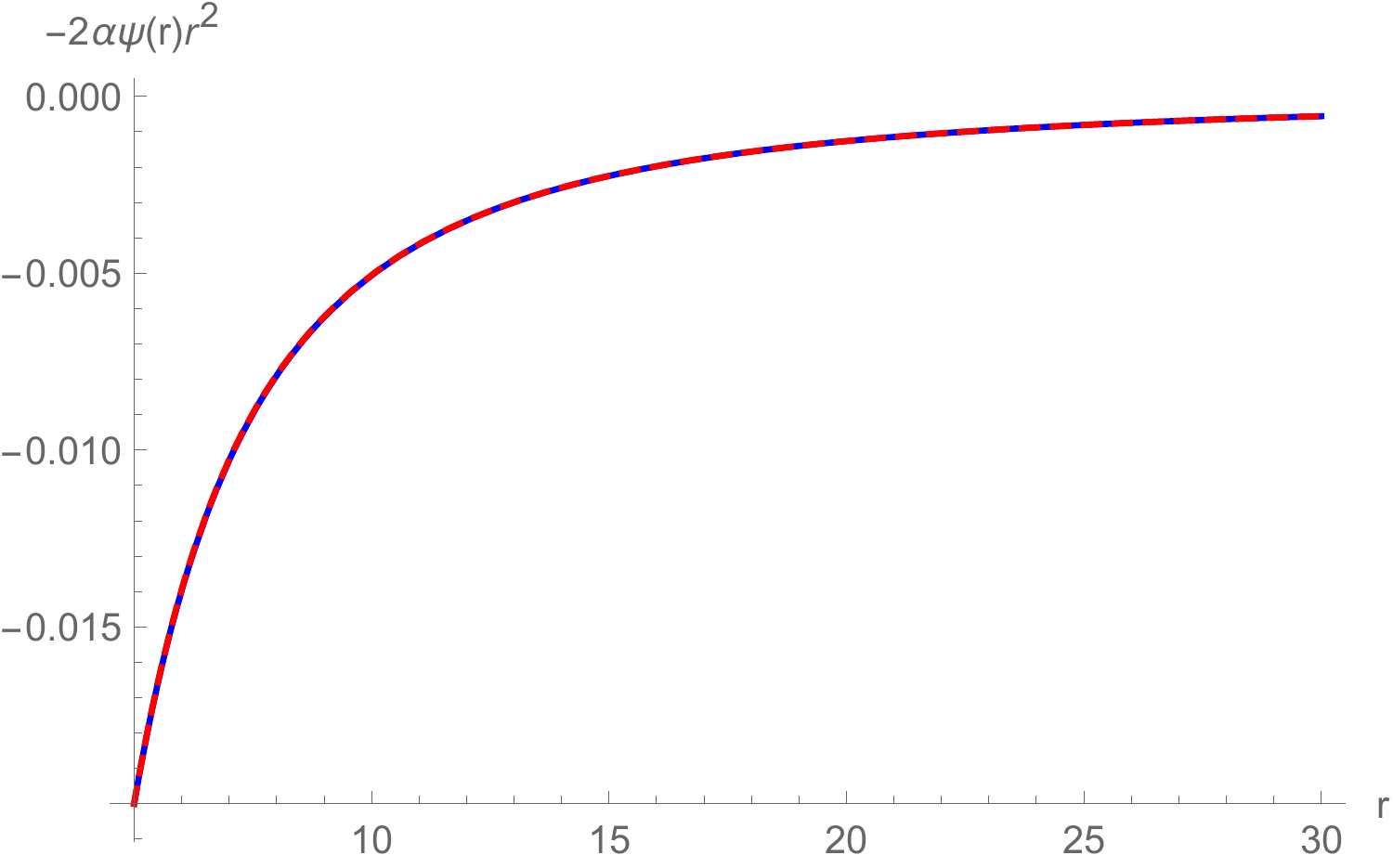}
\includegraphics[width=0.5\linewidth]{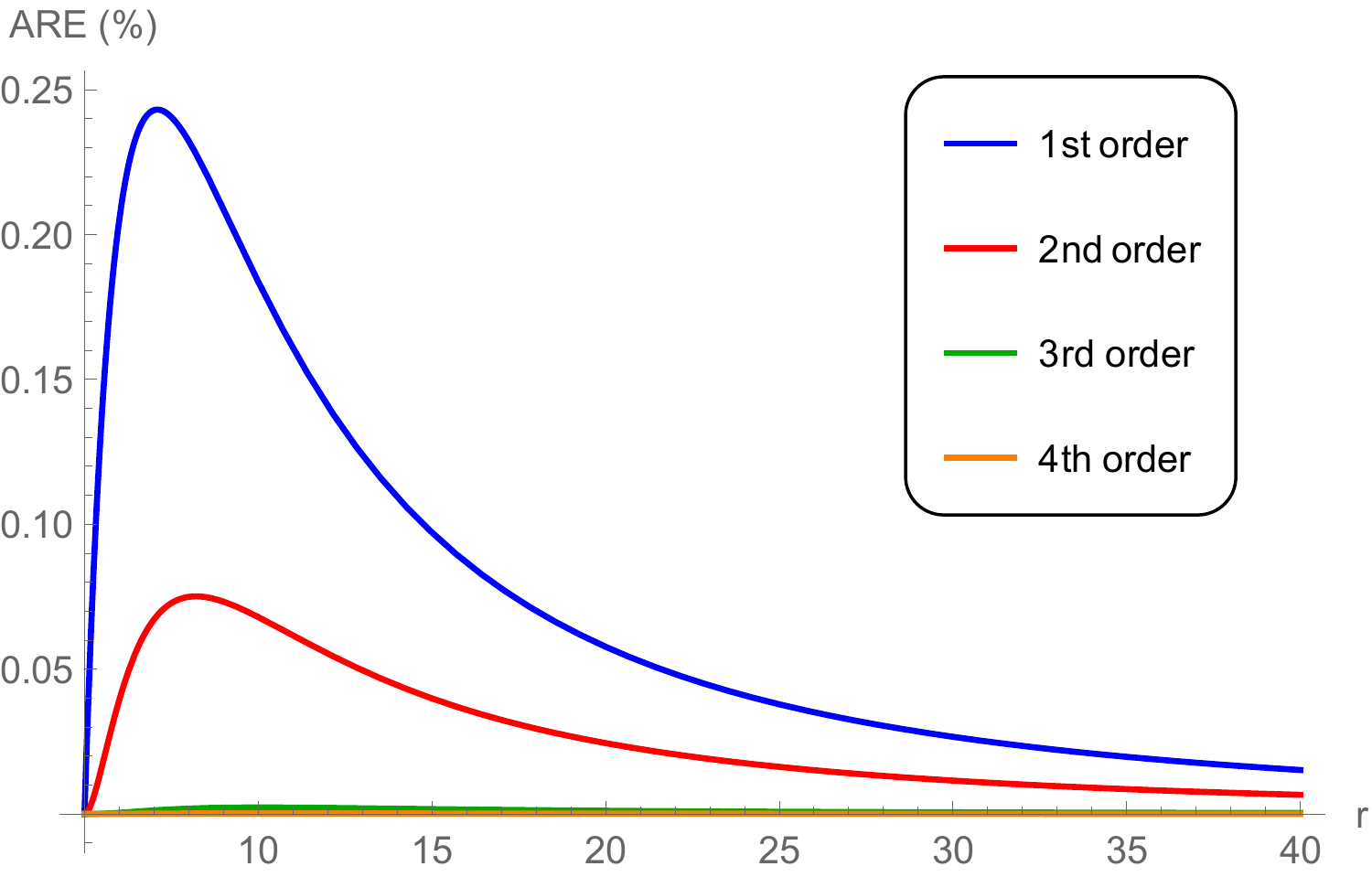}
\caption{The exact solution $-2 a\,\psi(r)\,r^2$ (blue curve) and its 1st order approximation (red-dashed curve) obtained via \eqref{OmegaCFA} (left panel). The absolute relative error in percentages ($\%$) for the first four orders of the approximation (right panel). The indicative values of the parameters used for these figures are $D=5,\,r_0=5,\,\mathcal{Q}=\frac{3}{2},\,\tilde {\alp}_2=\frac{1}{4},\,a=10^{-2}$\,.
}\label{fig:SR_D5}
\end{figure*}

In Fig.~\ref{fig:SR_D5} we fix $\theta=\pi/2$ and plot the ARE for the first four orders in the approximation \eqref{OmegaCFA} for the function $-2 a\,\psi(r)\,r^2$. It is clear that the parametrization scheme of Eq.~\eqref{OmegaCFA} provides an accurate description of the metric component already at the first order with a MARE smaller than $0.25 \%$ and also converges since the MARE is reduced significantly with every higher-order that is taken into account.

The fact that the approximated metric functions are impressively close to their exact expressions is a necessary but not sufficient condition to guaranty the accuracy of the HDCFA. To this end, in the next section we turn to the computation of the quasi-normal modes (QNMs) for black-hole solutions emerging in Einstein-Lovelock theory.

\Section{Quasinormal modes\label{sec:QNMs}}

\noindent The necessity of the parametrization we develop appears when solving various spectral problems, be it quasinormal modes, bound states, grey-body factors used for the estimation of intensity of Hawking radiation or others. Quasinormal modes and Hawking radiation are also important for higher-dimensional black holes, when considering various braneworld scenarios allowing for additional spacial dimensions \cite{Kanti:2004nr,Kanti:2005xa,Kanti:2009sn,Cardoso:2005vb}. When applying it to the Einstein-Lovelock theory with many coupling constants, constraining of many parameters of the metric from experiments would be an unfeasible problem. The representation of the black-hole metric in terms of only a few parameters would allow one to constrain the black-hole geometry in a simple way by imposing the limits on only those few parameters. Therefore, first of all, we need to understand how quickly the parametrization converges when considering spectral problems.

Thus, in order to further test the convergence of the parametrization, here we would like to calculate characteristic frequencies of oscillations, called \emph{quasinormal modes} of the Einstein-Lovelock black hole and compare them with those for the approximate metrics obtained by the parametrization at different orders.
This will give us an understanding on how practical the parametrization can be when used for spectral problems around higher-dimensional black holes.
The relatively simple illustration is to consider a test scalar field, which, unlike, the gravitational field, is governed by a much simpler effective potential. Although the wave equations for gravitational perturbations are well known, they have so lengthy and complex effective potentials that analysis of the Einstein-Lovelock black-hole's spectrum with many coupling constants would require considerable computer time.

The general covariant equation for a massless scalar field has the form
\begin{equation}\label{KGg}
\frac{1}{\sqrt{-g}}\partial_\mu \left(\sqrt{-g}g^{\mu \nu}\partial_\nu\Phi\right)=0\,,
\end{equation}
and after separation of the variables Eq.~(\ref{KGg}) takes the following general wavelike form:
\begin{equation}\label{wave-equation}
\frac{d^2\Psi}{dr_*^2}+\left(\omega^2-V(r)\right)\Psi=0\,.
\end{equation}
The "tortoise coordinate" $r_*$ is defined by the relation
$ dr_*=dr/f(r)$, and the effective potential is
\begin{equation}\label{scalarpotential}
V(r) = f(r)\left(\frac{\ell(\ell+D-3)}{r^2}+\frac{(D-4)(D-2) f(r)}{4 r^2} + \frac{D-2}{2 r}\frac{d f(r)}{dr}\right)\,.
\end{equation}
For an asymptotically flat black hole, quasinormal modes $\omega_{n}$ correspond to solutions of the master wave equation (\ref{wave-equation}) with the requirement of the purely outgoing waves at infinity and purely incoming waves at the event horizon:
\begin{equation}
\Psi_{s} \sim \pm e^{\pm i \omega r^{*}}, \quad r^{*} \rightarrow \pm \infty\,.
\end{equation}

In order to find quasinormal modes we shall use two independent methods:
\begin{enumerate}
\item the integration of the wave equation (before the introduction of the stationary ansatz, that is, with the second derivative in time instead of the $\omega^2$-term) in time domain at a given point in space \cite{Gundlach:1993tp}
\item the (Wentzel–Kramers–Brillouin) WKB method suggested by Will and Schutz \cite{Schutz:1985zz}, extended to higher orders in \cite{Iyer:1986np,Konoplya:2003ii,Matyjasek:2017psv} and combined with the usage of the Pade approximants in \cite{Matyjasek:2017psv}. We will use the 6th order WKB approach and use further the Pade approximants \cite{Matyjasek:2017psv} with $\tilde {m} =5$, where $\tilde {m}$ is defined in \cite{Konoplya:2019hlu}.
\end{enumerate}
Both methods are very well studied and applied in a large number of papers (see, for example, reviews \cite{Konoplya:2019hlu,Konoplya:2011qq}). Therefore, we will not describe them in detail here, but will simply show that both methods are in good agreement in the common parametric range of applicability.

\begin{figure*}[h!]
\includegraphics[width=0.5\linewidth]{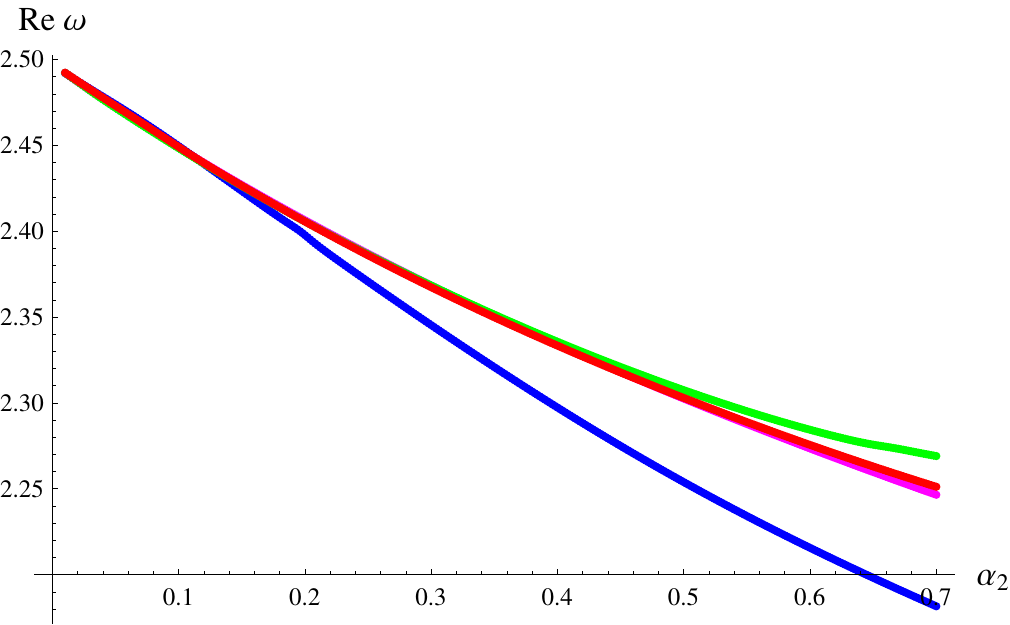}
\includegraphics[width=0.5\linewidth]{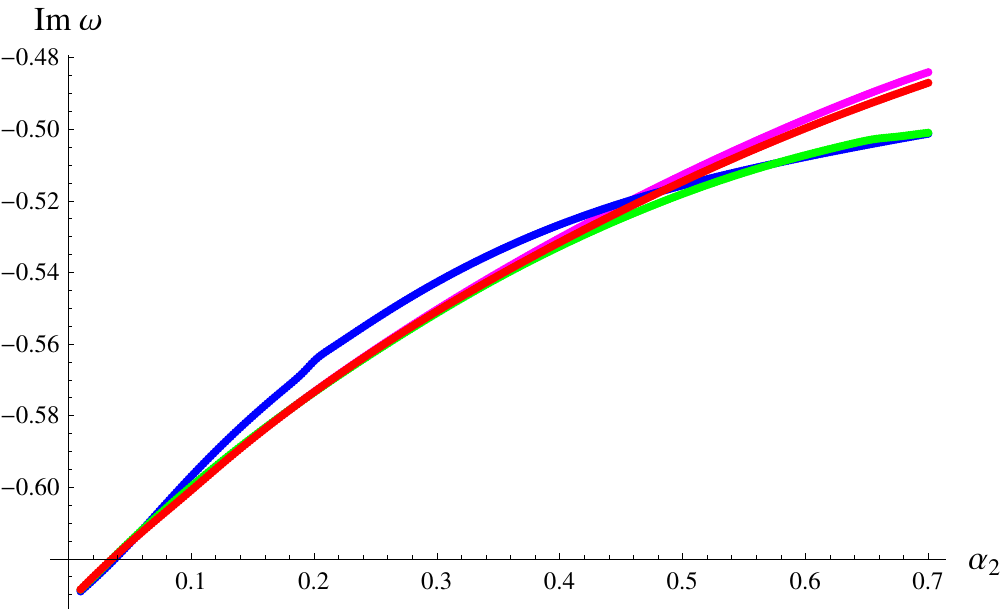}
\caption{The fundamental ($n=0$) quasinormal mode of a test scalar field as a function of $\tilde {\alpha}_2$ in the background of the Einstein-Gauss-Bonnet black hole for $\ell=2$, $D=7$, $Q=0$, $r_{0}=1$. The computations are done for the metric function approximated by the first (blue), second (green) and third (red) orders of continued fraction expansion. QNMs for the exact black-hole solution are given by the pink line. 
}\label{fig:QNM1}
\end{figure*}

\begin{figure*}[h!]
\includegraphics[width=0.5\linewidth]{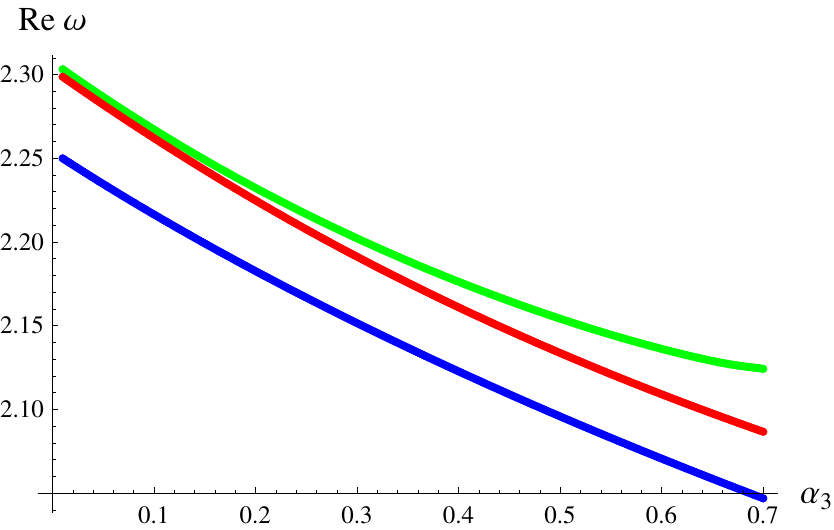}
\includegraphics[width=0.5\linewidth]{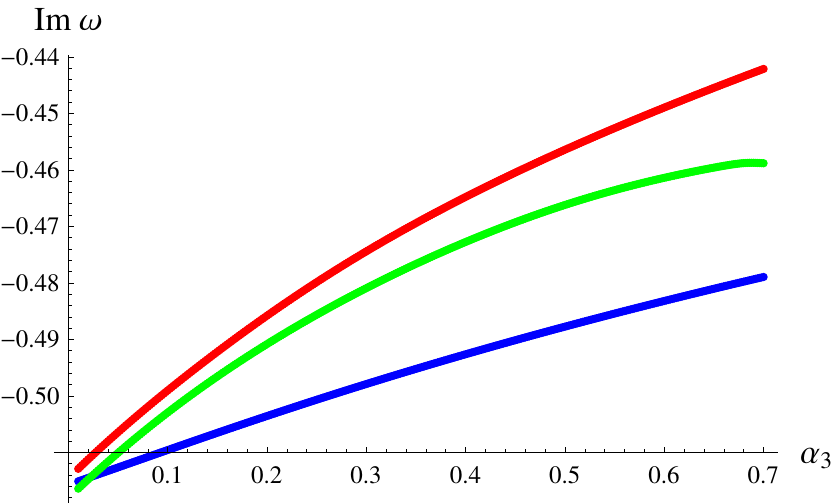}
\caption{The fundamental ($n=0$) quasinormal mode of a test scalar field as a function of $\tilde {\alpha}_3$ in the background of the Einstein-Lovelock black hole for $\ell=2$, $D=7$, $Q=0$, $\tilde {\alpha}_2 =0.5$, $r_{0}=1$. The computations are done for the metric function approximated by the first (blue), second (green) and third (red) orders of continued fraction expansion.
}\label{fig:QNM2}
\end{figure*}

\begin{table*}
\footnotesize	
\begin{tabular}{|c|c|c|c|c|c|c|}
  \hline
  % after \\: \hline or \cline{col1-col2} \cline{col3-col4} ...
  k & WKB ($\ell=0$) & T-d ($\ell=0$) & WKB ($\ell=1$) & T-d ($\ell=1$) & WKB ($\ell=2$) & T-d ($\ell=2$) \\
   \hline
  1 & $1.1301 - 0.5351 i$ & $1.1355 - 0.5365 i$ & $1.7049 - 0.5291 i$ & $1.7077 - 0.5277 i $ & $2.2755 - 0.5274 i$ & $2.2760 - 0.5258 i$ \\
   \hline
  2 & $1.1563 - 0.5382 i$ & $1.1548 - 0.5394 i $ & $1.7213 - 0.5283 i$ & $1.7227 - 0.5282 i$ & $2.2923 - 0.5239 i$ & $2.2922 - 0.5238 i$ \\
  \hline
  3 & $1.1581 - 0.5376 i$ & $1.1539 - 0.5386 i$ & $1.7215 - 0.5262 i$ & $1.7215 - 0.5272 i $ & $2.2909 - 0.5225  i$ & $2.2900 - 0.5230 i$\\
  \hline
  e & $1.1582 - 0.5376 i$ & $1.1541 - 0.5396 i$ & $1.7216 - 0.5261 i$ & $1.7218 - 0.5285 i$ & $2.2908 - 0.5224 i$ & $2.2910 - 0.5244 i$ \\
  \hline
\end{tabular}
\caption{The fundamental quasinormal mode ($n=0$) of a test scalar field for various values of the multipole number $\ell$ calculated with the help of the 6th order WKB method and using the Pad\'e approximants and time-domain (T-d) integration; $\tilde {\alpha}_2 =1/4$, $\tilde {\alpha}_3 = 1/2$, $\tilde {\alpha}_4 =1/7$, $\tilde {\alpha}_5 = 1/10$, $Q=0$, $r_{0}=1$, $k$ is the order of the continued fraction expansion in the parametrization.\label{Table2}}
\end{table*}

From Table~\ref{Table2} we see that indeed there is convergence to the quasinormal modes for the exact black-hole solution in the example for the Einstein-Lovelock theory with four coupling constants. There is a small discrepancy between the WKB and time-domain results related to worse accuracy of the WKB method for smaller multipoles $\ell$ and, at the same time, lack of a sufficiently long period of quasinormal oscillations for the $\ell=0$ case, so that the fitting of the time-domain profile by a sum of exponents with some excitation factors does not allow us to extract the frequencies with sufficient accuracy. Nevertheless, in both methods we see a clear convergence to the exact solution whose quasinormal modes are given in the last line of the Table~\ref{Table2}.

On the other hand, from Figs.~\ref{fig:QNM1}, \ref{fig:QNM2} we can see that for sufficiently large values of the coupling constant the deviation from the Tangherlini spectrum is larger and more orders of the continued-fraction expansion must be used to provide sufficient accuracy of the parametrization. In the case of the Einstein-Gauss-Bonnet solution~\eqref{GB_metric} (thus, with a single coupling constant $\tilde {\alpha}_2$) (see Fig.~\ref{fig:QNM1}) we reproduce the quasinormal modes obtained in \cite{Konoplya:2004xx,Abdalla:2005hu}. In the limit when all higher curvature corrections vanish, we have also reproduced quasinormal modes of the Tangherlini solution shown in Tables IV and V of \cite{Konoplya:2019hlu}.

In the above Figs.~\ref{fig:QNM1}, \ref{fig:QNM2} we can see that when the coupling constant is relatively small (about $\sim 0.2$), even the first order expansion provides sufficient accuracy of the parametrization, because the relative error in this case is evidently much smaller than the effect, that is, the deviation of the quasinormal frequency from its Tangherlini value. When the second-order parametrization is used it allows to describe even moderate values of the coupling constant ($\sim 0.7$). In general the concrete values of the relative error induced by the parametrization at each order depends on a number of factors which includes the number of coupling constants, their values and the number of spacetime dimensions $D$. Nevertheless, our general experience when calculated quasinormal modes can be summarized by the following statement: for relatively small deviations of the quasinormal frequencies from their Tangherlini limit the first-order approximation provides sufficient accuracy, while the second-order parametrization allows to test even moderate deviations from the Tangherlini geometry.

\Section{Conclusions\label{sec:Conclusions}}

\noindent In this work, we have introduced a parametrization scheme for the approximation of higher dimensional (HD), spherically symmetric and asymptotically flat black-hole metrics in an arbitrary metric theory of gravity. The method developed herein constitutes an extension of the Rezzolla-Zhidenko approach \cite{Rezzolla2014} that is mainly characterized by two features that set it apart from other Taylor-expansion based approximation methods. First, the parametrization is expressed in terms of a compact coordinate $x \equiv \left(r-r_0 \right)\,r^{-1}$ that increases monotonically from the radius $r_0$ of the outer event horizon of the black hole where $x=0$ up to asymptotic infinity $r \rightarrow \infty$ where $x=1$. Second, to encapsulate the features of the metric function close to the event horizon, a continued-fraction expansion (CFE) is introduced in terms of a tower of parameters [see Eqs.~\eqref{defAxBx}-~\eqref{TildeAB}].

As we have pointed out, the compact coordinate is modeled around the Schwarzschild metric function and this observation allowed us to extend the method to HD metrics. Inspired by the corresponding metric function of the Tangherlini solution \cite{Tangherlini1963}, we have introduced a new generalized compact coordinate as $\tilde {x}=1-\left(r_0/r \right)^{D-3}$ where $D$ is the total number of dimensions. Under this straightforward modification, we obtained a highly accurate parametrization technique for HD black-hole metrics that is valid everywhere in the spacetime outside (including) the event horizon up to asymptotic infinity.

A main difference between the four-dimensional and HD
continued-fraction approximations (CFAs) lies in the allowed values of the asymptotic parameter $a_0$ of the parametrization. By considering the asymptotic expansions of the metric functions at spatial infinity it is easy to associate this parameter with the asymptotic charge $\mathcal{Q}^2$ of the solution [see Eqs.~\eqref{HDasympt}]. In the context of 4D-CFA the resultant approximate metrics need to comply with stringent bounds set on the parameters by observations. In fact, $a_0$ can also be expressed as a combination of the PPN parameters $\beta$ and $\gamma$ \cite{Will2014a} the values of which are observationally determined. Taking into account these values, any metric obtained via the 4D-CFA formalism should have $a_0$ of $\mathcal{O}(10^{-4})$ \cite{Rezzolla2014}. On the other hand, no observational constraints exist for HD metrics and so $a_0$ in the HDCFA remains into play and should not be fixed \emph{$a$ priori} to $a_0=0$.

The extension of the parametrization to incorporate the description of rotating higher-dimensional black hole is a highly nontrivial task given the complexity of the problem since in $D$ extra dimensions the black hole can in principle rotate in $\lfloor (D-1)/2 \rfloor$ independent directions. Nevertheless, by restricting our investigation to the case of black holes with a single rotation parameter $a$, that corresponds to rotation on a two-plane on the brane, and imposing the slow-rotation condition $a\ll1$ we were able to formulate a parametrization scheme for HD black holes under these conditions.

As a first application of our method and in order to test its accuracy, we turned to the black-hole solutions that emerge in the context of the Einstein-Lovelock theory. We found that only the first few terms of the HDCFA  provide a very accurate approximation for the exact, albeit cumbersome expressions for the metrics in various dimensions and Lovelock curvature orders. Our investigation revealed that in the first order of the CFA, the maximum absolute relative error (MARE) between the exact and the approximate metric functions is smaller than $0.3 \%$ while for every consequent order of the approximation, the MARE is reduced by approximately one order of magnitude. The latter result serves as a test that emphatically verifies the convergence of our parametrization. We also provide our readers with a supplementary Mathematica\textregistered{} notebook where one can obtain for a given set of fixed values of the free parameters of the system the HDCFA expressions to the desired order.

The very small values of the MAREs mean that the Lovelock solutions for asymptotically flat black holes belong to the family of the so-called moderate metrics that require only first order in the CFA to yield expressions that deviate at most by a small fraction of $1\%$ from the exact metrics. To this end, we derived analytic expressions for the moderate metrics that are valid for arbitrary values of the free parameters and number of dimensions. The true merit of these expressions is their ultra-compact size contrary to the exact solutions that are quite lengthy and not very convenient to perform analytic computations with. In fact, the larger the number of dimensions the more dramatically the complexity of the exact solutions increases while our analytic approximations, at the cost of introducing an error of fractions of $1\%$, maintain their compact size.

Finally, in order to test the effectiveness of the parametrization for solutions of various spectral problems we calculated quasinormal modes of Einstein-Gauss-Bonnet and Einstein-Lovelock black holes. On various examples it is shown that when the deviation of the quasinormal modes from their Tangherlini values (corresponding to vanishing higher curvature corrections) are small, the first-order CFA usually provides sufficient accuracy, while when dealing with moderate values of the coupling constant and larger deviations from the  Tangherlini geometry, the second order CFA becomes necessary to keep the relative error at least one order smaller than the observable effect. This also shows that the parametrization can be effectively used to constrain the allowed black-hole geometry by constraining only a few parameters of the parametrization rather then dealing with multiple coupling constants of the Einstein-Lovelock black holes.

\vspace{0.5 cm}
\begin{acknowledgments}
The authors acknowledge the support of the grant 19-03950S of Czech Science Foundation (GA\v{C}R). This publication has been prepared with partial support of the ``RUDN University Program 5-100'' (R. K.).
\end{acknowledgments}

%\newpage
\bibliography{References}{}
\bibliographystyle{utphys}
\end{document}